# Econophysics of Macroeconomics: "Action-at-a-Distance" and Waves

Victor Olkhov
TVEL, Kashirskoe sh. 49, Moscow, 115409, Russia,
victor.olkhov@gmail.com

ABSTRACT

We present macroeconomic model that describes evolution of macroeconomic variables and macroeconomic waves on economic space. Risk ratings of economic agents play role of their coordinates on economic space. Aggregation of economic variables like Assets and Investment, Credits and Loans of economic agents at point *x* define corresponding macroeconomic variables as functions of time *t* and *coordinates x* on economic space. Evolution of macroeconomic variables is determined by economic and financial transactions between economic agents. Such transactions can occur between economic agents with any coordinates *x* and *y* and that reflect non-local "action-at-a-distance" character of internal macroeconomic interactions. For instance, Buy-Sell transactions between points *x* and *y* on economic space define dynamics of Assets at point *x* and Investment at point *y*. Aggregates of transactions between economic agents at point *x* and *y* on economic space define economic fields as functions of two coordinates. To describe dynamics of economic fields on economic space we derive hydrodynamic-like equations. For simple models of interactions between economic fields we derive hydrodynamic-like equations in a closed form and obtain wave equations for their perturbations. Economic field waves propagate on economic space and their amplitudes can grow up as exponent in time and may disturb economic stability. Diversities of macroeconomic and financial waves on economic space in simple models uncover importance of wave processes for macroeconomic modeling and forecasting.

# I. Introduction

This paper presents Econophysics approach to macroeconomics and develops parallels between description of multi particle systems and macroeconomic multi agent systems. Such approach is well known in Econophysics and there are many applications of methods of statistical physics, kinetics and hydrodynamics to description of economic systems [1-8]. As well there exist definite misunderstanding of Econophysics models by economic and financial researchers [9,10]. Indeed fundamental distinctions between economic and physical systems make direct applications of physical methods for economic modeling not too effective. Nevertheless certain parallels between economic and physical systems may allow use language of physics to develop economic theory on base of pure economic and financial issues.

We develop model of macroeconomics that is based on pure economic notions with help of some parallels between economics and physics. The core issues of our approach concern introduction of economic space [11-13] that allow describe multi agent economic systems alike to multi particle systems in physics. For decades international rating agencies estimate risk ratings of corporations and banks. These risk ratings help take Investment decisions, determine credit policy and avoid losses due to default. We propose that risk assessment methodologies and practice can be developed in such a way that can provide risk ratings for all economic agents of economic and finance system. Such "simple" assumption requires tough efforts and development of risk methodologies and risk econometrics. It requires cooperation of risk agencies and market authorities, banking and financial regulators, businesses and Government Statistical Bureaus, academic and business researchers, etc. Let assume that all econometric and regulatory problems can be solved and risk ratings can be provided on a regular basis for all economic agents: for huge banks and international corporations and for small companies and households. How that can help for economic modeling?

We propose that risk ratings of economic agents allow establish first "parallel" between economics and physics – introduce coordinates of economic agents [11-13]. Risk ratings take values of risk grades that can be treated as points of certain discreet space. Let treat risk ratings of economic agents as their coordinates on such a space. That distributes all economic agents by their coordinates and develops parallels to description of particles in physics. Let assume that risk assessment methodology can be extended and define continuous risk grades like points on *R*. Then one can plot economic agents by their risk ratings as coordinates on continuous space. Let call such a space determined by risk grades as economic space and define it's dimension by number of risks measured simultaneously.

Distribution of numerous economic agents by their risk ratings as coordinates on economic space presents a model *alike to* multi-particle systems in physics. There are at least two distinctions that underline vital differences between economic and physical systems. Almost all classical physics problems are described by systems of particles those have only few properties like mass and electric charge. Contrary to that, economic agents or economic particles as we call them further, have enormous number of economic and financial variables like Capital and Value, Demand and Supply, Assets and Debts, Investment and Credits, Production Function and Wages and etc. Aggregations of economic and financial variables define macroeconomic and macro financial variables. Modeling relations between numerous macroeconomic variables define subject of economic theory. Economic particles (economic agents) have tens of economic and financial variables and that increase variety and complexity of economic system modeling to compare with description of multi-particle systems in physics.

Second essential distinction between physical and economic systems concern type of interactions between particles in physics and interaction between economic agents. Physics is based on local type of interactions – particle 1 interacts with particle 2 when coordinates are nearly the same. Economic and financial interactions are completely different. For example, economic agent with coordinates *x* can Sell goods, attract Investment, Buy resources, receive Loans from economic agents with coordinates *y* on economic space. Thus economic agent with coordinate *x* on economic space can carry out economic or financial transaction with economic agent with any coordinate *y* on economic space. Thus economics presents wonderful example of "action-at-a-distance" interactions between agents on economic space but we shall not argue any "parallels to" Mach's principle and etc. We simply highlight complete distinctions between economic and physical systems and present economic model in "terms" that are alike to kinetics and hydrodynamics. In [11-13] for simplicity we presented macroeconomic models that describe dynamics of macroeconomic variables in the assumption that transactions between economic agents are *local*. In other words we assumed that economic transactions between agents on economic space occurs for agents with nearly same coordinates only. Such simplifications allow use parallels to *local* interaction between physical particles and describe interaction between macroeconomic variables by simple differential operators [13]. In this paper we develop macroeconomic model that takes into account non-local "action-at-a-distance" transactions between agents on economic space.

It is wonderful that economic theory already has certain model of *"action-at-a-distance"* interactions between economic agents. Nearly eighty ears ago famous economist Wassily Leontief developed Input-Output Analysis or inter-industry Tables framework [14-18]. In his Nobel Lecture Leontief [16] expressed key issue of input-output framework as: "Direct interdependence between two processes arises whenever the output of one becomes an input of the other: coal, the output of the coal mining industry, is an input of the electric power generating sector". Thus Leontief's inter-industry Tables describe *"action-at-a-distance"* transactions between economic agents those belong to different industries. In this paper we suggest a simple thing: let substitute distributions of economic agents by industries with distributions of economic agents by their risk ratings as points of economic space. That can allow use Leontief's ideas presented by inter-industry Tables and develop a model of *"action-at-a-distance"* transactions between economic agents that have different coordinates on economic space. Description of economic and financial transactions between agents on economic space gives description of corresponding macroeconomic variables. Thus we can obtain model of evolution of macroeconomic variables determined by *"action-at-a-distance"* transactions between economic agents. To describe dynamics of these transactions we derive hydrodynamic-like equations. These equations permit describe interactions between different economic and financial transactions and for simple models of such interaction derive hydrodynamic-like equations in a closed form. For such simple models of interaction between economic transactions we derive wave equations on their perturbations. Dynamics of any complex systems should be accompanied by complex wave generations, propagations and interactions. Our model presents evidence that macroeconomics and finance also governed by a wide range of macroeconomic wave processes on economic space.

We present pure theoretical treatment of macroeconomics. Up now no econometric data that can verify or reject predictions of our macroeconomic model exist. Current risk assessment data provided by risk rating agencies are not sufficient to develop macroeconomic models on economic space. We do hope that enhancement of risk assessments procedures, econometric observations and data performance can improve economic modeling, forecasting and management.

The rest of the paper is organized as follows. In Section II we introduce of *n*-dimensional economic space. In Section III we reconsider Leontief's input-output framework and introduce economic fields that describe transactions between economic agents at different points on economic space. In Section IV we describe economic fields by hydrodynamic-like equations. In Section V for simple two *conjugate* Credits-Loans and Payments-on-Credits economic fields interaction model we derive hydrodynamic-like equations in a closed form. In Section VI we derive wave equations on economic field disturbances. Conclusions are in Section VII.

## II. Definition Of Economic Space

Description of multi-agent system alike to multi-particle system requires economic analogy of physical space that allows define coordinates of economic agents similar to coordinates of physical particles. We suggest use risk ratings of economic agents as their coordinates on economic space. Definition of economic space was presented in [11-13] but we repeat it briefly here for convenience.

International rating agencies [19-21] estimate risk ratings of economic agents as Banks and Corporations, Firms and Enterprises. Risk ratings take values of risk grades and noted as *AAA, BB, C* and so on. Let treat risk grades like *AAA, BB, C* as points $x_1, x_2,.. x_m$ of discreet space. Let propose, that risk assessments methodologies can estimate risk ratings for all agents of entire economics: for huge Banks and for small households. That will distribute all economic agents of the entire economics over points of finite discreet space determined by set of risk grades. There are a lot of different risks those impact economic processes. Let regard grades of single risk as points of one-dimensional space and simultaneous rating assessments of *n* different risks as measurements of coordinates of economic agent on *n*-dimensional space. Let propose, that risk assessments methodologies can be generalized in such a way that risk grades can fill continuous space $R$. Thus risk grades of $n$ different risks establish $R^n$.

Let define economic space as any mathematical space that map economic agents by their risk ratings as space coordinates. Number of risks ratings measured simultaneously determines dimension of economic space. Let put positive direction along each risk axis as risk growth direction. Let assume that all economic agents of entire economics are "independent" and thus sum of extensive (additive) economic variables of any subset of agents equals economic variable of the entire subset. For example, sum of Assets of any two "independent" economic agents equal their collective Assets. Let assume that econometric data provide info about risk ratings and economic variables of each economic agent. These assumptions require significant development of current econometrics and statistics. Quality, accuracy and granularity of current U.S. National Income and Product Accounts system [22] give hope that all these problems can be solved.

It is obvious that there exist a lot of different economic and financial risks those affect evolution of economic agents. It is impossible to take into account all possible risks. To develop reasonable economic theory one should select some major risks and neglect minor risks. Definition of economic space $R^n$ requires selection of $n$ risks with major impact on economic agents and macroeconomic processes. These $n$ risks define initial state of economic space $R^n$. Selection of most valuable risks requires procedures that allow measure and compare influence of different risks on entire economics and economic agents. Assessment and comparison of different risks and their influence on economic agents establish tough problems and such models should be developed. Risk assessments methodologies and procedures, comparison of risk influence on performance of economic agents and on macroeconomic dynamics can establish procedures alike to physical measurement theory. It may help develop relations between economic theory on economic space, econometric data and economic statistics similar to interdependence between theory and measurements in physics. Solution of this problem requires close collaboration between physicists and economists.

Economic and financial risks have random nature and can unexpectedly arise and then vanish. Thus some current risks that define initial representation of economic space $R^n$ can accidentally disappear and other different risks may come to play. Thus economic space representation can be changed randomly. Description of economic dynamics and financial forecasting for time term $T$ requires prediction of $m$ main risks that can play major role in a particular time term and can define economic space $R^m$. Such set of $m$ risks determine target state of economic space $R^m$. Transition of economic modeling on initial economic space $R^n$ to target economic space $R^m$ requires description of decline of action of initial set of $n$ risks on entire economics and description of growth of influence of new $m$ risks. Such stochastic scenarios are completely different from physical models that study complex dynamics of random fields and particles determined on constant physical space.

Current macroeconomics describes relations between macroeconomic variables as Demand and Supply, Production Function and Investment, Economic Growth and Consumption each treated as function of time. Introduction of economic space gives ground for definition of macroeconomic variables as functions of time and *coordinates*. This small step opens doors for wide application of mathematical physics methods and models that should be transformed to adopt economic and financial phenomena's.

Below we present macroeconomic model on economic space $R^n$ in the assumption that economic agents are under action of constant set of $n$ risks. We describe macroeconomics alike to kinetics and hydrodynamics and derive hydrodynamic-like and wave-like equations. Up now notions of waves in economics and finance are used to describe Kondratieff waves [23], inflation waves, crisis waves, etc. All these issues don't describe any waves but time oscillations of economic variables only. Description of waves requires space. Introduction of economic space gives ground for development of economic wave theory and establishes ground for unified approach to various economic problems. Economic space gives new look on option pricing theory and allows describe Markov processes on economic space [11,12]. In this paper we develop macroeconomic model with "action-at-a-distance" interactions between economic agents on economic space. For brevity let further call economic agents as economic particles or e-particles and economic space as e-space.

## III. Definition of Economic Fields

Let regard economics as system of e-particles. Economic variables of e-particles are determined by transactions between e-particles. Let call transactions between e-particle *i* with coordinates *x* and e-particle *j* with coordinate *y* as economic field $a_{ij}(x,y)$. These economic fields describe exchanges of Resources, Commodities, Energy, Assets, Profits, Liabilities etc. between e-particles. Eighty years ago W. Leontief [14] developed input-output analysis that describe inter-industry resources exchange. Let consider core issue of his approach: "the output of one becomes an input of the other" as any economic or financial transactions between e-particles on e-space.

Each e-particle has many economic and financial variables like Demand and Supply, Assets and Liabilities, Credits and Loans, etc. Leontief's ideas allow define economic fields between any two e-particles. Let call economic or financial variables of two e-particles as *mutual* if "the output of one becomes an input of the other". For example, Credits as output of Banks are *mutual* to Loans as input of Borrowers. Assets are output of Investors are *mutual* to Liabilities as input of Debtors. Let call any exchange between e-particles by *mutual* economic or financial variables as action of corresponding economic field between them. Different *mutual* economic or financial variables interact by different economic fields. Let define that economic field $a_{1,2}(x,y)$ between e-particle *1* at point *x* and e-particle *2* at point *y* describes exchange by economic variables $B_{in}(1,x)$ and $B_{out}(2,y)$ at moment *t*. Let $a_{1,2}(x,y)$ be output economic variable $B_{out}(2)$ from e-particle *2* to e-particle *1* and $a_{1,2}(x,y)$ input of economic variable $B_{in}(1)$ of e-particle *1* from e-particle *2* at moment *t*. So, $a_{1,2}(x,y)$ describes speed of change of economic variable $B_{in}(1,x)$ of e-particle *1* due to exchange with e-particle *2* and $a_{1,2}(x,y)$ describes speed of change of economic variable $B_{out}(2,y)$ of e-particle *2* due to exchange with e-particle *1*. Then economic variable $B_{in}(1,x;y)$ of e-particle *1* changes due to action of economic field $a_{1,2}(x,y)$ with e-particles at point *y* as follows:

$$dB_{in}(1,\mathbf{x};\mathbf{y}) = \sum_i a_{1,i}(\mathbf{x},\mathbf{y})\,dt \qquad (1.1)$$

and vice versa

$$dB_{out}(2,\mathbf{x};\mathbf{y}) = \sum_i a_{i,2}(\mathbf{x},\mathbf{y})\,dt \qquad (1.2)$$

For example such relations may describe change of Credits (output) from e-particle *2* to e-particle *1* due to Credits-Loans economic field between them. For such case $B_{in}(1)$ equals total Loans received by e-particle *1* and $B_{out}(2)$ equals total Credits issued by e-particle *2*. Sum of economic field over all input e-particles equals speed of change of output economic variable $B_{out}(2)$ of economic particle *2*. Let assume that all extensive economic and financial variables of economic particles can be presented as pairs of *mutual* economic variables or can be describes by *mutual* variables. For example Value of e-particle (Value of Corporation or Bank) don't take part in transactions but is determined by pairs of *mutual* economic variables like Credits and Loans, Sales and Purchases, Assets and Liabilities, etc. Let assume that all extensive economic variables can be described by Eq.(1.1,1.2) or through other *mutual* economic variables. Thus we assume that economic fields describe dynamics of all extensive economic variables of e-particles and hence describe dynamics and evolution of macroeconomics and finance.

# IV. Hydrodynamic-Like Model Of Economic Fields

This Section describes economic fields of e-particles on e-space alike to kinetics and hydrodynamics [24,25]. Let assume that economic fields between e-particle at point *x* and e-particle at point *y* are determined by exchange of *mutual* economic variables like Assets and Liabilities, Credits and Loans, Buy and Sell, Demand and Supply and so on. Different economic fields describe action between different *mutual* economic variables. For example, if e-particle "one" at point *x* gets Loan (input) of amount *cl* from e-particle "two" at point *y* then e-particle "two" at point *y* at time *t* provides a Credit (output) of amount *cl (Credit-Loans)* to e-particle "one" at point *x*. Thus function $cl_{i,j}(t,x,y)$ on *2n*-dimensional e-space describes *Credit-Loans* economic field or *Credit-Loans* transaction between e-particle *i* at points *x* and e-particle *j* at point *y*. That is similar to treatment of economic variable *cl* of e-particles on *2n*-dimensional e-space and allows develop parallels to kinetics and hydrodynamics. Let study *Credit-Loans* (*CL*) economic field model.

## A. Macroeconomic variables

Macroeconomics describes interdependence between macroeconomic variables like GDP, Assets, Credits, Taxes, Consumption ant etc. Each macroeconomic variable is composed of economic variables of economic agents (e-particles). For example, macroeconomic Assets are determined as sum of Assets of all "independent" e-particles. Introduction of e-space allows define macroeconomic variables as functions of time and coordinates. Thus mutual interdependence between macroeconomic variables can be described as relations between functions on e-space with help of methods that have parallels to mathematical and statistical physics. We introduced e-space densities of macroeconomic variables in [13] and give it briefly here for convenience. Let study economics that consist of numerous e-particles on e-space $R^n$. Let assume that economics is under action of $n$ major risks and each e-particle on e-space $R^n$ at moment $t$ is described by coordinates $x=(x_1,...x_n)$ and velocities $v=(v_1,...v_n)$. Let describe economics that has $l$ macroeconomic variables and each "independent" e-particle has $l$ economic variables $(u_1,...u_l)$. Let assume that values of economic variables equal $u=(u_{1i},...u_{li})$, $i=1,..N(x)$. Each extensive economic variable $u_j$ at point $x$ defines macroeconomic variable $U_j$ as sum of economic variables $u_{ji}$ of $N(x)$ "independent" e-particles at point $x$

$$U_j = \sum_i u_{ji} \; ; \quad j = 1,..l; \quad i = 1,...N(x)$$

For each macroeconomic variable $U_j$ let define analogy of impulses $P_j$ as

$$\boldsymbol{P}_j = \sum_i u_{ji} \boldsymbol{v_i} \; ; \quad j = 1,..l; \quad i = 1,...N(x)$$

Let follow [24] and introduce economic distribution function $f=f(t,x;U_1,..U_l, P_1,..P_l)$ on $n$-dimensional e-space that determine probability to observe macroeconomic variables $U_j$ and impulses $P_j$ at point $x$ at time $t$. $U_j$ and $P_j$ are determined by corresponding values of e-particles that have coordinates $x$ at time $t$. Averaging of $U_j$ and $P_j$ within distribution function $f$ allows establish transition of macroeconomic description from kinetic-like approximation that takes into account economic variables of separate e-particles to hydrodynamic-like approximation that neglects e-particles granularity. Let define macroeconomic density function $U_j(t,x)$

$$U_j(t,\boldsymbol{x}) = \int U_j \, f(t,\boldsymbol{x},U_1,...U_l,\boldsymbol{P}_1,..\boldsymbol{P}_l) \, dU_1...dU_l d\boldsymbol{P}_1..d\boldsymbol{P}_l \qquad (2.1)$$

and impulse density $P_j(t,x)$ as

$$\boldsymbol{P}_j(t,\boldsymbol{x}) = \int \boldsymbol{P}_j \, f(t,\boldsymbol{x},U_1,...U_l,P_1,..P_l) \, dU_1...dU_l d\boldsymbol{P}_1..d\boldsymbol{P}_l \qquad (2.2)$$

That allows define e-space velocity $v_j(t,x)$ of macroeconomic density $U_j(t,x)$ as

$$U_j(t,\boldsymbol{x})\boldsymbol{v}_j(t,\boldsymbol{x}) = \boldsymbol{P}_j(t,\boldsymbol{x}) \qquad (2.3)$$

Densities $U_j(t,x)$ and impulses $P_j(t,x)$ are determined as mean values of aggregates of corresponding economic variables of separate e-particles with coordinates $x$. Functions $U_j(t,x)$ can describe macroeconomic e-space density of Demand and Supply, Assets and Debts, Production Function and Value Added and so on. Usage of distribution function $f=f(t,x;U_1,..U_l, P_1,..P_l)$ allows describe any statistical moments of macroeconomic variables like $<U_j^m>$, correlations between economic variables $<U_j U_i>$ and so on. Operators $<..>$ define averaging by distribution function $f$. Evolution of macroeconomic variables and their densities is determined by transactions between e-particles. For example macroeconomic Credits *C(t,y)* at point *y* are determined by transactions of Credits that are provided from e-particles at point *y* to e-particles at point *x*. Paper [13] describes relations between macroeconomic densities for *local* approximation that takes into account only transactions between e-particles with same coordinates near point *x*. In other words we describe transactions between e-particles with nearly same risk ratings only. Such simplification is *alike to* local interaction between physical particles and allow describe interaction between macroeconomic variables *U(t,x)* and V(t,*x*) on e-space by differential operators [13].

Real economic processes are determined by economic and financial transactions between e-particles with any different coordinates *x* and *y* – any different risk ratings *x* and *y*. In other words e-particles with coordinate *x* (risk rating *x*) can Buy-Sell, Credits-Loans, Investment-Liabilities and etc., with e-particles with coordinate *y*. Such "action-at-a-distance" transactions between e-particles at points *x* and *y* describe complex interdependence between macroeconomic variables *U(t,x)* and V(t,*y*) that can not be described by differential operators as in [13]. To describe relations between macroeconomic variables in the approximation that takes into account "action-at-a-distance" properties of transactions between e-particles on e-space one should define macroeconomic transactions densities – economic fields – alike to Eq.[2.1-2.3].

B. Macroeconomic transactions and economic fields

Let use Credits-Loans transactions as example to describe macroeconomic transactions alike to description of macroeconomic variables. Let assume that at moment *t* there are *N(x)* e-particles at point *x* and *N(y)* e-particles at point *y*. Let state that velocities of e-particles at point *x* equal $\upsilon=(\upsilon_1,…\upsilon_{N(x)})$. Let state that each of *N(y)* e-particles numbered as *j=1,…N(y)* at point *y* have Credit-Loans transactions equal $cl_{i,j}(x,y)$, *i=1,…N(x)*, *j=1,…N(y)* with e-particles numbered as *i=1,…N(x)* at point *x* at moment *t*. In other words, if e-particle *i* at point *x* increases it's Loans by $cl_{i,j}(x,y)$ due to Credits from e-particle *j* at point *y* at moment *t* then e-particle particle *j* at point *y* provides Credit $cl_{i,j}(x,y)$ at moment *t* to e-particle *i* at point *x*. Let state that $cl_{i,j}(x,y)$ define Credit-Loans economic field between e-particles at points *x* and *y*. That exactly reproduce Leontief's framework for Credit-Loans exchange between two e-particles. Let assume that e-particles on e-space are "independent" so, that at moment $t$ e-particles at point $x$ receive Loans $l_j(x,y)$ from e-particle $j$ at point $y$ on e-space $R^n$

$$l_j(x,y) = \sum_i cl_{ij}(x,y); \quad i = 1, ... N(x)$$

Thus $l_j(x,y)$ equals total rise of Loans of all e-particles at point $x$ due to Credits from e-particle $j$ at point $y$ at moment $t$. Sum $c_i(x,y)$

$$c_i(x,y) = \sum_j cl_{ij}(x,y); \quad j = 1, ... N(y)$$

equals total rise of Credits $c_i(x,y)$ provided by all e-particles at point $y$ to e-particle $i$ at point $x$ at moment $t$. $cl(x,y)$ defined as

$$cl(x,y) = \sum_{ij} cl_{ij}(x,y); \quad i = 1, ... N(x); i = 1, ... N(y) \quad (2.4)$$

equals growth of Credits provided from all e-particles at point $y$ to e-particles at point $x$ and equals rise of Loans of all e-particles at point $x$ due to Credits from all e-particles at point $y$ at moment $t$. These relations introduce Credits-Loans transactions as economic fields between points $x$ and $y$ very similar to input-output inter-industry tables. We replace output from *industry 1* by output from all e-particles at point $y$ and input of *industry 2* by input of all e-particles at point $x$. Our approach allows replace inter-industry tables by description of functions of two variables $(x,y)$ on $n$-dimensional e-space. Eq.(2.4) define macroeconomic Credit-Loans economic field $cl_{ij}$ that has parallels to density function. To develop parallels to kinetics and hydrodynamics let introduce velocities of economic fields. For economic field $cl_{ij}$ let define *impulses $p$* $=(p_X, p_Y)$ alike to impulses of e-particles [11-13]:

$$p_X = \sum_{i,j} cl_{ij} v_i; \quad i = 1, ... N(x); j = 1, ... N(y) \quad (2.5)$$

$$p_Y = \sum_{i,j} cl_{ij} v_j; \quad i = 1, ... N(x); j = 1, ... N(y) \quad (2.6)$$

Credits-Loans economic field $cl(x,y)$ between two points $x$ and $y$ on e-space is defined as function of two variables on e-space $R^n$ and takes random values due to random motion of e-particles. Let use averaging procedure to obtain regular value of economic field between two points $x$ and $y$ that is *alike to* transition from kinetic approximation to hydrodynamic approximation in physics.

Let follow [24] and define economic distribution function $f=f(t, z=(x,y); cl, p=(p_X,p_Y))$ on $2n$-dimensional e-space $R^{2n}$ that determine probability to observe Credits-Loans economic field $cl$ at point $z=(x, y)$ with impulses $p =(p_X, p_Y)$ at time $t$. Economic field $cl$ and impulses $p=(p_X, p_Y)$ are determined by Eq.(2.4-2.6). Economic field takes random values at point $z=(x,y)$ and averaging of *cl* within distribution function *f* allows establish transition from economic kinetic-like approximation that takes into account transactions between separate e-particles to economic hydrodynamic-like approximation that determine "mean" economic fields as functions of $z=(x,y)$. Let define Credits-Loans economic field density function *CL(z=(x,y))* as

$$CL(t, \mathbf{z} = (\mathbf{x}, \mathbf{y})) = \int cl\, f(t, \mathbf{x}, \mathbf{y}; cl, \mathbf{p}_X, \mathbf{p}_Y)\, dcl\, d\mathbf{p}_X\, d\mathbf{p}_Y \qquad (3.1)$$

Let introduce impulse densities $P(t,x,y)=(P_X(t,x,y), P_Y(t,x,y))$ as

$$\mathbf{P}_X(t, \mathbf{z} = (\mathbf{x}, \mathbf{y})) = \int \mathbf{p}_X\, f(t, \mathbf{x}, \mathbf{y}; cl, \mathbf{p}_X, \mathbf{p}_X)\, dcl d\mathbf{p}_X d\mathbf{p}_Y \qquad (3.2)$$

$$\mathbf{P}_Y(t, \mathbf{z} = (\mathbf{x}, \mathbf{y})) = \int \mathbf{p}_Y\, f(t, \mathbf{x}, \mathbf{y}; cl, \mathbf{p}_X, \mathbf{p}_Y)\, dcl d\mathbf{p}_X d\mathbf{p}_Y \qquad (3.3)$$

and define e-space velocity $v(t,z=(x,y))=(v_X(t,z),v_Y(t,z))$ of economic field *CL(t, z)* as:

$$\mathbf{P}_X(t, \mathbf{z}) = CL(t, \mathbf{z}) \mathbf{v}_X(t, \mathbf{z}) \qquad (3.4)$$

$$\mathbf{P}_Y(t, \mathbf{z}) = CL(t, \mathbf{z}) \mathbf{v}_Y(t, \mathbf{z}) \qquad (3.5)$$

Economic fields may describe many important economic properties. For example, Assets-Liabilities economic field *AL(z=(x,y))* describes distribution of speed of Assets allocations from point *y* as function of *x*. For given *x* economic field *AL(z=(x,y))* describes speed of Liabilities changes at point *x* as function of *y* on e-space $R^n$. Due to Eq.(1.1-1.2) integral of economic field density *AL(x,y)* by *y* over e-space $R^n$ defines speed of change of total Liabilities at point *x*. Integral by *x* over e-space $R^n$ determines speed of change of total Assets at point *y*. Similar distributions can be obtained for Credits-Loans economic field *CL(x,y)*. It describes distributions of Credits provided from point *y* to point *x* at moment *t*. Such economic fields allow describe transactions by *mutual* economic and financial variables like Assets and Liabilities, Credits and Loans etc., and study other economic problems. For example economic fields can describe dynamics of maximum of Credits allocations at point *x=x(t)* on e-space or distance between position of maximum source of Credits and position of maximum Loan borrowers on e-space. Economic fields like *CL(t,x,y)* are functions of two variables *x* and *y* on e-space $R^n$. Usage of economic fields as functions of two variables on e-space for macroeconomic modeling is complementary to input-output inter-industry tables framework developed by Leontief. We simply apply key Leontief's ideas on input-output framework to describe relations between e-particles (economic agents) on e-space and thus enlarge usage of Leontief's method. That enhances description of input-output inter-industry tables by mathematical physics methods for modeling economic fields as functions on e-space. Below we derive hydrodynamic-like equations to describe Credits-Loans *CL(t, z=(x,y))* economic field model.

## C. Hydrodynamic-Like Approximation of Economic Fields

Economic field *CL(t,x,y)* and impulses *P(t,x,y)* are determined by Eq.(3.1-3.5) as mean values of transactions between e-particles at point *x* and *y*. Economic fields similar to *CL(t,z=(x,y))* can describe *mutual* variables as Credits and Loans, Buy and Sell, Demand and Supply and so on. That allow regard economic fields on *2n*-dimensional e-space $R^{2n}$ similar to economic variables densities on e-space $R^n$ [11-13] and alike to mass density distribution *ρ(t,x)* in physical kinetics [24]. We use term "alike" to underline vital differences between nature of physical kinetics and hydrodynamics on one hand and properties of economic variables and economic fields on the other hand. We state absence of direct analogies between description of economic and physical systems. Nevertheless we derive hydrodynamic-like equations on economic fields.

Let define economic field *A(x,y)* between two *mutual* economic variables $A_{in}(x)$ and $A_{out}(y)$ on e-space $R^n$. *A(x,y)* equals output $A_{out}(y)$ from point *y* and equals input $A_{in}(x)$ at point *x* at moment t. Functions *A(t,z=(x,y))* and $v(t,z=(x,y))=(v_X(t,z)),v_Y(t,z)))$ are determined on *2n*-dimensional e-space $R^{2n}$. Let regard economic field *A(t,z)* as economic density [11-13] on *2n*-dimensional e-space $R^{2n}$. Continuous Equations (4.1) and Equations of Motion (4.2) on *A(t,z)* take form:

$$\frac{\partial A}{\partial t} + div(\boldsymbol{v}A) = Q_1 \qquad (4.1)$$

$$A\left[\frac{\partial \boldsymbol{v}}{\partial t} + (\boldsymbol{v}\cdot\nabla)\boldsymbol{v}\right] = \boldsymbol{Q}_2 \qquad (4.2)$$

Left side of Eq.(4.1) describes the flux of density *A(t,z)* through surface of unit volume on e-space $R^{2n}$ and $Q_1$ describes factors that change *A(t,z)*. Economic field *A(t,z)* can change in time and during motion of the selected volume on e-space due to economic reasons. Left side of Equation of Motion describes flux of impulse density *P(t,z) = A(t,z)v(t,x)* through unit volume surface on e-space $R^{2n}$. If one takes into account Continuity Equation then for simplicity left side of Equation of Motion takes (4.2.) $\boldsymbol{Q}_2$ describes factors that change density *A(t,z)* and velocity *v (t,z)*.

## D. Economic Field Model

Economic fields describe interactions between economic agents (e-particles) on e-space. In [11,13] we describe models of mutual dependence of economic variables on e-space. These models were based on assumptions that economic variables of e-particles at point *x* depend on *conjugate* economic variables at *same point x* of e-space. We assumed *local* transactions between e-particles and hence *local* interaction of macroeconomic variables of e-particles on e-space. It is obvious that assumption on *local* transactions between e-particles or transactions between e-particles with same coordinates only simplifies macroeconomic model and neglect economic and financial transactions between e-particles with different coordinates. Nevertheless such simplification allows develop "simple" description of macroeconomic variables and derive wave equations on e-space for variables disturbances. Even such simplified model discovers extreme diversity and complexity of possible behavior of macroeconomic variables on e-space [11,13].

Meanwhile, real economic relations are much more complex. E-particles at point *x* on e-space can exchange by economic and financial transactions with e-particles at any other point *y* on e-space. Thus economic variables of e-particles and macroeconomic variables of entire economics at point *x* depend on *conjugate* economic variables at different points on e-space. This is alike to action of certain *economic fields* that describe transactions of economic variables between points *x* and *y* on e-space. For example, economic field *CL(t,x,y)* between points *x* and *y* describes Credits provided from e-particles at point *y* to e-particles at point *x* at moment *t*. Such economic field *CL(t,x,y)* describes how Loans at point *x* depend on Credits from point *y* on e-space at moment *t*.

To derive economic hydrodynamic-like equations on economic fields in a closed form let study interaction between economic fields. Let define *conjugate* economic fields alike to *conjugate* economic variables [11,13]. Let state that economic field *A(t,z)* depend on other economic fields or economic variables that are different form *mutual* variables that define *A(t,z)*. Let denote economic field *B(t,z)* as *conjugate* to economic field *A(z)* if *B(z)* or their velocities determine right hand side factors $Q_1$ and $Q_2$ of hydrodynamic-like Eq.(4.1- 4.2). Any economic fields can have one, two or many *conjugate* economic fields that determine right hand side of Eq.(4.1-4.2). For example, Credits-Loans economic field may depend on Payments-on-Credits economic field, Demand-on-Investment, Buy-Sell transactions that require additional funds and etc. Thus economic field can depend on many *conjugate* economic fields. Below we present simples model of interaction between economic fields that describes interaction between two *conjugate* economic fields - Credits-Loans and Payments-on-Credits and derive hydrodynamic-like equations on economic fields in a closed form.

## V. Two Conjugate Economic Fields Model

To derive Eq.(4.1-4.2) in a closed form let study simple model of mutual dependence between two *conjugate* economic fields as Credits-Loans *CL(z)* and Payments-on-Credits *PC(z)*. Let define Payments-on-Credits *PC(z=(x,y))* economic field as all pay offs to e-particles at point *y* that are made by e-particles from point *x* due to Loans they received from e-particles at point *y*. Thus Payments-on-Credits economic field *PC(z=(x,y))* describes income flow of e-particles at point *y* received at moment *t* from e-particles at point *x*. Economic field *CL(z=(x,y))* describes Credits provided from e-particles at point *y* to e-particles at point *x* at moment *t*. Credits-Loans *CL(z)* and Payments-on-Credits *PC(z=(x,y))* economic fields describe important properties of macroeconomics and finance. These economic fields are responsible for economic growth and financial sustainability and their descriptions are extremely complex. Introduction of e-space allows establish and study various models that describe interactions between economic variables and economic fields and different approximations of real economic and financial processes.

Let start with simple model and assume that Credits-Loans economic field *CL(t,z)* at moment *t* depends on Payments-on-Credits economic field *PC(t,z=(x,y))* at moment *t* only. Our assumption means that Creditors, like Banks at point *y* take decisions to provide Credits to point *x* on base of Payments-on-Credits from point *x* to point *y* at moment *t*. Such assumption simplifies the problem but allows develop reasonable model of their mutual interdependence.

To describe evolution of Credits-Loans field *CL(t,z)* let define factors $Q_1$ and $\mathbf{Q}_2$ of Eq.(4.1-4.2). Let assume that $Q_1$ on the right hand side of Continuity Equation (4.1) for Credits-Loans field *CL(t,z)* is proportional to divergence of Payments-on-Credits velocity *u(z)* on *2n*-dimensional e-space $R^{2n}$:

$$Q_1 \sim PC(\mathbf{z})\nabla \cdot \mathbf{u}(\mathbf{z}) \qquad (5.1)$$

Positive divergence of Payments-on-Credits *PC(z)* economic field velocity *u(t,z)* describes growth of flux of Payments-on-Credits and that may attract Creditors at point *y* to increase their Credits at point *x*. Negative divergence of velocity *u(t,z)* means that Payments-on-Credits *PC(z)* flow decrease and that may prevent Creditors at point *y* from providing additional Credits to point *x*. Let assume that $Q_1$ factor that defines right hand side of Continuity Equation (4.1) for Payments-on-Credits field *PC(t,z)* is proportional to divergence of Credits-Loans velocity *v(t,z)*:

$$Q_1 \sim CL(\mathbf{z})\nabla \cdot \mathbf{v}(\mathbf{z}) \tag{5.2}$$

Positive divergence of Credits-Loans *CL(z)* field velocity *v(t,z)* describes growth of Credits-Loans flux and that may increase Payments-on-Credits *PC(t,z)*: growth of Credits from point *y* to point *x* of e-space may induce growth of Payments-on-Credits from *x* to *y*. As well negative divergence of Credits-Loans *CL(z)* flux describes decline of Credits flow from *y* to *x* and that may reduce Payments-on-Credits from *x* to *y*. It is obvious that we neglect time gap between providing Credits and Payments-on-Credits and other economic and financial factors that determine decisions on providing Credits to simplify the model. Let determine $Q_2$ factors in Equations of Motion for Credits-Loans economic field *CL(z=(x,y))* Eq.(4.2). Let assume that velocity *v(t,z)* of Credits-Loans field *CL* depends on right hand side factor $Q_2$ that is proportional to gradient of Payments-on-Credits *PC(t,z)*:

$$\mathbf{Q}_2 \sim \nabla PC(\mathbf{z}) \tag{5.3}$$

These relations propose that Credits-Loans economic field velocity *v(t,z)* grows in direction of higher Payments-on-Credits. Let make same assumptions on $Q_2$ that determines Equation of Motion (4.2) for Payments-on-Credits field velocity *u(t,z)*:

$$\mathbf{Q}_2 \sim \nabla CL(\mathbf{z}) \tag{5.4}$$

Payments-on-Credits field velocity *u(t,z)* grows up in the direction of higher Credits-Loans. Assumptions (5.1-5.4) give hydrodynamic-like equation on two *conjugate* economic fields Credits-Loans and Payments-on-Credits in a closed form. Continuity Equations:

$$\frac{\partial CL}{\partial t} + \nabla \cdot (\mathbf{v}CL) = a_2 PC(\mathbf{z})\nabla \cdot \mathbf{u}(\mathbf{z}) \tag{6.1}$$

$$\frac{\partial PC}{\partial t} + \nabla \cdot (\mathbf{u}PC) = a_1 CL(\mathbf{z})\nabla \cdot \mathbf{v}(\mathbf{z}) \tag{6.2}$$

Equations of Motion:

$$CL(\mathbf{z})\left[\frac{\partial \mathbf{v}}{\partial t} + \mathbf{v} \cdot \nabla \mathbf{v}\right] = b_2 \nabla PC(\mathbf{z}) \tag{6.3}$$

$$PC(\mathbf{z})\left[\frac{\partial \mathbf{u}}{\partial t} + \mathbf{u} \cdot \nabla \mathbf{u}\right] = b_1 \nabla CL(\mathbf{z}) \tag{6.4}$$

Eq.(6.1-6.4) give ground for derivation of wave equations on disturbances of *conjugate* economic field disturbances.

## VI. Economic Field Wave Equations

Let derive equations on disturbances of economic fields in linear approximation. Let simplify the problem and assume

$$CL(\mathbf{z}) = CL + cl(\mathbf{z}) \,; \, PC(\mathbf{z}) = PC + pc(\mathbf{z}) \tag{7.1}$$

Let assume that *CL* and *PC* are constant or their changes are negligible to compare with variations of small disturbances *cl(z), pc(z), v(z)* and *u(z)* and let neglect nonlinear factors in Eq.(6.1-6.4). These assumptions allow derive equation on economic fields disturbances in linear approximation *alike to* derivation of acoustic wave equations [12] in fluids. Continuity Equations:

$$\frac{\partial cl}{\partial t} + CL \nabla \cdot \mathbf{v} = \alpha_2 PC \nabla \cdot \mathbf{u} \,; \quad \frac{\partial pc}{\partial t} + PC \nabla \cdot \mathbf{u} = \alpha_1 CL \nabla \cdot \mathbf{v} \tag{7.2}$$

Equations of Motion:

$$CL \frac{\partial \mathbf{v}}{\partial t} = \beta_2 \nabla \, pc(\mathbf{z}) \,; \quad PC \frac{\partial \mathbf{u}}{\partial t} = \beta_1 \nabla \, cl(\mathbf{z}) \tag{7.3}$$

These equations allow derive equations on *cl* and *pc*

$$[\frac{\partial^4}{\partial t^4} - a\Delta \frac{\partial^2}{\partial t^2} + b\Delta^2] cl(t, \mathbf{z}) = 0 \tag{7.4}$$

$$a = \alpha_1 \beta_2 + \alpha_2 \beta_1 \,; \, b = \beta_1 \beta_2 (\alpha_1 \alpha_2 - 1)$$

Derivation of Eq.(7.4) from Eq.(7.2-7.3) is simple and we omit it. For

$$c_{1,2}^2 = \frac{a +/- \sqrt{a^2 - 4b}}{2} > 0$$

Eq.(7.4) take form of bi-wave equations:

$$(\frac{\partial^2}{\partial t^2} - c_1^2 \Delta)(\frac{\partial^2}{\partial t^2} - c_2^2 \Delta) cl(t, \mathbf{z}) = 0 \tag{7.5}$$

Here $c_{1,2}$ - different speeds of economic field disturbances waves propagating through e-space. Green function of bi-wave equation (7.5) equals convolution of Green functions of common wave equations with wave speeds equal $c_1$ and $c_2$. Thus even simple δ-function source induce complex wave response. Eq.(7.4) or Eq.(7.5) validate diversity of wave processes that govern interactions of economic fields. Economic field disturbances can induce waves that propagate through macroeconomic domain on e-space and may cause time fluctuations of macroeconomic variables as GDP, Investments, Profits, Demand, Supply, etc.

Let show how simple economic field waves can impact macroeconomic properties and cause time oscillations of macroeconomic variables. Let take simple solution for economic field waves of Eq.(7.4) as:

$$cl(t, \mathbf{z}) = \cos(\mathbf{k} \cdot \mathbf{z} - \omega t) \exp(\gamma t) ; \quad \mathbf{k} = (k_x, k_x) \tag{8.1}$$

$$\omega^2 = k^2 \frac{\sqrt{4b+3a^2}+2a}{8} > 0 \; ; \; \gamma^2 = k^2 \frac{\sqrt{4b+3a^2}-2a}{8} > 0 \tag{8.2}$$

Amplitudes of simple harmonic waves (8.1) of Credits-Loans economic field disturbances *cl(t,z)* for *γ>0* grow as *exp(γt)*. Relations (8.1) describe propagation of Credits-Loans field disturbances wave *cl(t,z)* in the direction of wave vector **k** with frequency *ω* determined by (8.2). Integral by coordinate *x* for Credits-Loans field *CL(t,z=(x,y))* determines distribution of rate of providing Credits *C(t,y)* from point *y* at moment *t*. Integral by coordinate *y* for *C(t,y)* at time *t* on e-space determines value of all Credits *C(t)* provided in macroeconomics at moment *t*. Value of Credits *C(t)* provided at moment *t* define rate of macroeconomic activity. Decline of rate of Credits *C(t)* at moment *t* reflect economic recession. Thus integral of (8.1) by e-space coordinates over macroeconomic domain reflect phases of economic growth or recession.

Due to definition of e-space in Section 2 coordinates of e-particles reflect their risk ratings. Thus, for simplest *1*-dimensional e-space *R* Credits-Loans economic field *CL(t,z=(x,y))* is determined on e-space $R^2$. Let assume that risk ratings of e-particles are reduced by minimum $X_{min}$ and maximum $X_{max}$ risk grades. For simplicity let take $X_{min}=0$ and $X_{max}= X$ that define macroeconomic domain as

$$0 \le x \le X \tag{8.3}$$

Due to (7.1) Credits-Loans economic field *CL(t,z=(x,y))* is presented as

$$CL(t, \mathbf{z}) = CL + cl(t, \mathbf{z}) \tag{8.4}$$

and *CL* is constant or its variations are small to compare with variations of disturbances *cl(t,z)*. For assumption (8.3) integral *C(t)* of (8.1; 8.4) by *x* and *y* on e-space $R^2$ gives

$$C(t) = C_0 + c(t) \; ; \; C_0 \sim CL\, X^2 \tag{8.5}$$

$$c(t) = -\frac{4exp(\gamma t)}{k_x k_y} \sin\left(\omega t - \frac{k_x+k_y}{2}X\right) \sin\frac{k_x}{2}X \sin\frac{k_y}{2}X \tag{8.6}$$

Hence speed of Credits *C(t)* provided in macroeconomics at moment *t* follows time oscillations with frequency *ω* and can grow up as *exp(γt)* for *γ>0*. Exponential growth of Credits-Loans disturbances will perturb other macroeconomic variables and that may reflect "overheating" of economic growth. On the other hand exponential growth of Credits-Loans wave amplitude will violate applicability of initial model (7.1-7.5) and that require additional considerations. On the other hand linear Eq.(7.5) for given wave speed *c* may have wave solutions with different wave vectors ***k*** and hence with different frequencies *ω*. Hence waves with random wave vectors ***k*** may induce random oscillations of Credits disturbances provided in macroeconomics at moment *t*. This conclusion establishes relations between time oscillations of macroeconomic Credits and time oscillations of economic growth on one hand and simple model of Credits-Loans economic field waves on e-space on the other hand. In other words, relations between macroeconomic variables like Growth, Credits, Investment, Assets etc., treated as functions of time can be determined by complex interaction between *conjugate* economic fields on e-space. Equations on economic field disturbances can admit wave solutions with amplitudes as *exp(γt)* and such waves on e-space can cause irregular time fluctuations of macroeconomic variables. We present only simplest consequences of economic fields interaction models to show examples of diversity and complexity of hidden macroeconomic processes on e-space.

## VII. Conclusions

Introduction of e-space establishes certain parallels between description of economic and physical systems and opens wide opportunities for application of statistical and mathematical physics methods for macroeconomic modeling. As well strong distinctions between nature of physical and economic systems requires complete reformulation of common of models but allow development of kinetic-like and hydrodynamic-like macroeconomic models.

All macroeconomic variables are determined by corresponding variables of economic agents. Economic and financial transactions between economic agents define interactions and interdependence between macroeconomic variables. Economic and financial transactions can occur between economic agents with different risk ratings or different coordinates ***x*** and ***y*** on economic space. That outlines "action-at-a-distance" character of transactions on economic space. It is great that certain model of such non-local transactions was developed by W.Leontief nearly 80 years ago and is known in economics as input-output inter-industry tables framework. Leontief's ideas allow develop macroeconomic model that describe "action-at-a-distance" interaction between macroeconomic variables on economic space. We just replace allocation of economic agents by industries with allocation of economic agents by their risk ratings or points of economic space. That permits replace inter-industry tables by economic fields on economic space. Such substitution allows describe transactions between different points on economic space by hydrodynamic-like equations. Interactions between *conjugate* economic fields can be described by *local* hydrodynamic-like equations on economic space. To show advantages of such approach to macroeconomic modeling we derive hydrodynamic-like equations in a closed form for interactions between Credits-Loans and Payment-on-Credits economic fields. These equations admit derivation of wave equations for economic field disturbances alike to derivation of acoustic equations in hydrodynamics. Description of generation, propagation and interaction of economic field waves can be important for forecasting of macroeconomic variables and can help for macroeconomic policy-making. Macroeconomic variables as Demand and Supply, Investment and GDP, Assets and Liabilities and etc., can be described as functions on economic space and corresponding economic fields describe their evolution. Modeling and forecasting of macroeconomic dynamics requires development of economic field models on economic space. Complexity of macroeconomic relations and diversity of economic processes that are described by economic fields interactions on economic space leave few chances for adequate modeling by "mainstream" general equilibrium, DSGE, decision making, game theories and etc. [26-30]. Further development of macroeconomic and economic fields interactions models on economic space requires close collaboration of physicists and economists.

Economic space notion is a core issue of our approach. Introduction of economic space as generalization of risk ratings of economic agents permit describe economic agents by their coordinates on economic space. Economic agents can move on economic space alike to particles and that cause changes of macroeconomic variables. Such interpretation helps develop bridge between description of macroeconomics as set of economic agents and description of multi-particles systems. Nature of economic phenomena is completely different from physical phenomena. Economic space is determined by set of most valuable risks and has different representations for different set or major risks. Random properties of risk nature cause random changes of economic space representation. Thus it seems impossible to develop deterministic macroeconomic forecast as random nature of risks growth and decline insert permanent uncertainty into macro dynamics and modeling on economic space. Possibility to measure and select most valuable risks may establish procedure to validate the initial and target set of risks and to prove or disprove initial model assumptions. It makes possible to compare predictions of economic and financial models with observations and helps outline causes of disagreement between theoretical predictions and macroeconomic reality.

Development of macroeconomic models on economic space need appropriate econometric foundations that are absent now. Our model is pure theoretical as there are no econometric data that can verify or reject predictions of our theory. Verification requires development of risk assessment of economic agents and collecting data on economic and financial transactions between them. Risk assessment methodologies should be extended to provide risk ratings for all economic agents, for huge corporations and banks with billions dollars assets and for small householders with few thousands dollars income. Risk assessment methodologies should allow introduce risk grades – points of economic space - that can establish discreet space or continuous space $R$ as well. That requires collective efforts of Central Banks and Economic Regulators, Rating Agencies and Market Authorities, Businesses and Government Statistical Bureaus, Academic and Business Researchers, etc. Econometric foundations that are required for studies of Leontief's input-output inter-industry tables were successfully established. We do hope that problems required for macroeconomic modeling on economic space can be solved also and that diversity and complexity of problems aroused on economic space may be interesting for physicists.

## Acknowledgements

This research did not receive any specific grant from TVEL or funding agencies in the public, commercial, or not-for-profit sectors and was performed on my own account.